\documentclass[12pt]{article}
\usepackage{newtxtext,newtxmath}
\usepackage{graphicx,lineno}
\usepackage[letterpaper,margin=1in]{geometry}

\renewenvironment{abstract}
	{\quotation}
	{\endquotation}

\date{}

\makeatletter
\renewcommand{\fnum@figure}{\textbf{Figure \thefigure}}
\renewcommand{\fnum@table}{\textbf{Table \thetable}}
\makeatother

\usepackage{scicite}
\usepackage{url}

\def\scititle{
    Lightest and heaviest nuclei as keys to the origin of ultrahigh energy cosmic rays
}
\begin{document}

\title{\bfseries \boldmath \scititle}


\author{
	Anna~M.~Suliga$^{1\ast\dagger}$,
	Glennys~R.~Farrar$^{1\ast\dagger}$,\and
	\small$^{1}$Center for Cosmology and Particle Physics, Department of Physics,\and
    \small{New York University, New York, NY 10003, USA}\and
	\small$^\ast$Corresponding authors. Emails: a.suliga@nyu.edu and gf25@nyu.edu\and
	\small$^\dagger$These authors contributed equally to this work.
}

\maketitle

\begin{abstract} \bfseries \boldmath
The origin of ultrahigh energy cosmic rays (UHECRs) has been a long-standing mystery.   A unique feature of UHECRs originating in binary neutron star mergers (BNSMs) is two distinct populations of light elements -- those directly accelerated in the jets, and photo-disintegration products of massive nuclei accelerated in the bulk outflow. We show that Pierre Auger Observatory spectrum and composition data confirm the existence of the predicted two populations at $\geq \!7.2\sigma$ significance, having rigidity ($R\equiv E/Ze$) cutoffs agreeing with prior estimates. The data also favor 5-10\% of UHECRs originating as r-process nuclei, a signature of a BNSM origin.  Our results provide strong evidence that UHECRs are produced by merging neutron stars, and show how UHECRs and UHE-neutrinos elucidate otherwise inaccessible aspects of neutron star mergers. 
\end{abstract}

\noindent
Ultra-High Energy Cosmic Rays (UHECRs) are subatomic particles that have been accelerated to extraordinarily high energy -- millions of times higher than is achieved in CERN's Large Hadron Collider.  Discovering the nature and origin of UHECRs has been a major scientific challenge since their first detection more than 60 years ago~\cite{Linsley}.  Leading source candidates have been Active Galactic Nuclei (AGN)\cite{hillas84}, formed by the accretion of mass onto supermassive black holes, long Gamma Ray Bursts  (GRBs)~\cite{waxman95} following the collapse of massive stars, Tidal Disruption Events (TDEs)\cite{fg08}, when a star is torn apart in the gravitational field of a supermassive black hole, and cluster accretion shocks~\cite{blandford+ClusterAccel18}. 
The most recent proposal, which we test in this paper, is that UHECRs are produced in the merger of two neutron stars~\cite{fBNS-prl25,fBNSApJL25}.  

We analyze the latest Pierre Auger Observatory observations~\cite{PierreAuger:2025eun, PierreAuger:2021hun, PierreAuger:2026qbt, augerDNNPRD25} of the UHECR spectrum above $10^{17.5}\;$eV and the composition above $10^{17.7}\;$eV, and find evidence of two distinctive predictions of the binary neutron star merger (BNSM) scenario for UHECR production: \\ 1. The presence of a secondary higher energy jet component of p and/or He, in addition to the protons and alpha particles that arise as photo-disintegration products of the more abundant higher mass (but lower energy-per-nucleon) cosmic rays produced in the magnetized outflow. 
\\ 2. The possible presence of a component of UHECRs heavier than iron.

We also calculate the spectrum of UHE neutrinos produced by UHECR interactions as they escape the ejecta, and show how this UHE neutrino spectrum encodes the redshift history of binary neutron star mergers.

\subsection*{Overview}
\label{sec:Overwiev}

Studying UHECRs is difficult because they are so rare. Their flux is less than than one per square kilometer per century, but nonetheless it has been established that they are atomic nuclei~\cite{augerXmaxMeas14,augerXmaxInterp14}.  Until recently, the tools for discriminating between different source options have been limited.  
Helpfully, the distance to the sources of such high energy events is less than $\sim 100$ Mpc, thanks to the Greisen Zatsepin Kuzmin (GZK) horizon~\cite{Zatsepin:1966jv, Greisen:1966jv}, but tracing even the highest energy events back to their sources is imprecise due to uncertainties in their deflections in the magnetic field of our Galaxy~\cite{uf23,uf26arxiv}.  Several lines of evidence point to UHECRs being produced in powerful transient events in average galaxies~\cite{bf23,ufAma23}. However, observing a temporal correlation between a UHECR and its transient source is impossible because a UHECR's arrival is delayed by thousands of years due to its non-rectilinear propagation in extragalactic magnetic fields.  

An important clue to the source of UHECRs is the observed tight correlation between a UHECR's charge and its energy.   The spectrum of a single source naturally depends on rigidity ($R=E/Ze$), since a particle's deflection in a magnetic field is determined by its rigidity.  However a simple ``Peters Cycle" behavior for the \emph{ensemble} of UHECR sources was shown in~\cite{EhlertOikonomouUnger23} to be incompatible with the population diversity of the then-leading source candidates: AGN, long GRBs, and TDEs.  
This ``standard source" requirement led one of us to the proposal~\cite{fBNS-prl25} that UHECRs are accelerated in the magnetized outflow of binary neutron star mergers. 
The masses of known BNS systems are the same to within $\approx 25$\%~\cite{Ozel:2016oaf,LIGOScientific:2020aai} and the post-merger magnetic field is generated by a gravitationally-driven dynamo~\cite{Kiuchi24}, so the narrow mass range of neutron star binaries translates to highly similar acceleration conditions across the BNSM population.

Reference~\cite{fBNSApJL25} used results in the literature on nucleosynthesis in a BNS merger and the initial conditions on the magnetic fields from~\cite{Kiuchi24}, to develop a first-approximation model for the BNSM production of UHECRs.  There are two distinctive regions in the BNSM ejecta:  the largest magnetic fields are in polar jets of 10-12$^\circ$ opening angle, while a magnetized turbulent outflow fills the rest of the solid angle.  The field strength in the jet is an order-of-magnitude greater than in the outflow~\cite{Kiuchi24}, enabling it to accelerate to higher energies~\cite{fBNSApJL25}.  The polar region contains protons or helium with their relative abundance depending on the duration of a possible short-lived hypermassive neutron star prior to collapse, see, e.g.,~\cite{Perego+LightElems22}.  
At larger angles, nucleosynthesis rapidly assembles elements with 
proton and neutron numbers $Z,N \geq 20$.  The heaviest elements, which are created in a succession of rapid neutron captures (r-process), are preferentially formed in the equatorial region~\cite{radice+BNSnucleosyn22}~\footnote{The relative abundance of different UHECR nuclei has not yet been predicted, so not finding evidence for $A>56$ in the data would not undermine the BNSM scenario -- but the extreme rarity of r-process elements outside of BNS merger ejecta, means observing UHECRs with $A>56$ would be near-definitive evidence in favor of the BNSM scenario. The very highest energy Fly's Eye ``OMG" and TA's Amaterasu events are difficult to explain except if they have $Z>26$, but due to energy uncertainties and small statistics these events are not definitive evidence of r-process UHECRs~\cite{fBNS-prl25}.}. 

As the highly magnetized plasma expands, a small fraction of the nuclei get caught up in the acceleration process.  They reach their maximum energy about a day after the merger, when the synchrotron loss timescale becomes larger than the acceleration timescale~\cite{fBNSApJL25}.  Thanks to particle-in-cell (PIC) simulations for magnetized turbulence~\cite{cfm24}, the radius at which this occurs can be estimated to be about $\sim 10^{14}$ cm~\cite{fBNSApJL25},  leading to a cutoff rigidity ($R \equiv E/Ze$) of 5-7 EV~\cite{fBNSApJL25}, in agreement with observation~\cite{augerCombFitJCAP23}.  Initially, there is no p or He  in the outflow outside the jets, but these are produced through photo-dissociation and inherit the energy per nucleon of their progenitors.  Therefore the cutoff energy of the bulk light component is about 3 EeV/nucleon.  For the p or He in the jets, where the acceleration mechanism may be diffusive shock acceleration (DSA), we have no PIC simulations to fix the cutoff given the magnetic field, but based on the order-of-magnitude higher magnetic field strength in the jet, the cutoff energy in the jet should be at least 10 EeV for protons or 35 EeV for helium~\cite{fBNSApJL25}.

\subsection*{Modeling Cosmic Ray Sources}
\label{sec:Modeling Cosmic Ray Sources}

The analysis presented in our work builds on the Unger-Farrar-Anchordoqui (UFA) modeling of the UHECR flux~\cite{Unger:2015laa} which was further explored and extended in~\cite{Muzio:2019leu, Muzio:2021zud, Muzio:2022bak}.
The essence of the UFA treatment is inclusion of interactions of the UHECR nuclei with the photon field surrounding the CR accelerator, as well as with the extragalactic background light such as the CMB, en route to Earth. The photodisintegration while escaping the source picks off nucleons and naturally leads to the appearance of the ``ankle" feature in the observed spectrum of the UHECRs, and also explains the specific energy-dependence of the light nuclei at energies around and below the ankle~\cite{PierreAuger:2025eun,augerMixed16}. 

The minimal set of parameters describing the source of UHECRs is: the power law index of the accelerated CRs, $\gamma_\mathrm{inj}$; the fractions, $f_A$, of different nuclei emerging from the accelerator at a fixed energy in the power-law portion of the spectra; and the rigidity cut-off $R_\mathrm{max}$, which we take here to be the same for all $A$ -- a more detailed treatment is outlined in~\cite{fBNSApJL25}.
A key attribute of the environment surrounding the source is the mean number of interactions, $\langle N_\mathrm{int}\rangle$, of a fiducial nucleus (Fe, at 10 EeV) while escaping the source environment; $\langle N_\mathrm{int}\rangle$ is equal to the ratio of the escape time to the interaction time, denoted $R_{19}^{Fe}$ by~\cite{Unger:2015laa,Muzio:2019leu} and $r_\mathrm{esc}$~by~\cite{Muzio:2021zud}.  We take the radiation surrounding the source  to be a black-body of temperature $T$.  Two more parameters characterize the rigidity dependence of the escape of the UHECRs from the source environment: the ratio of the source size $L$ to the coherence length $\lambda_c$ of the magnetic field, $r_\mathrm{size}\equiv L/\lambda_c$, and the characteristic rigidity scale of UHECR diffusion in the source $R_\mathrm{diff} \equiv (\lambda_c / 6.9\;\mathrm{kpc}) (B/\mathrm{\mu G})\;\mathrm{EV}$. The latter two parameters were introduced in~\cite{Muzio:2021zud} to more accurately describe the energy dependence of the diffusion as measured in the diffusion simulations of~\cite{Globus:2007bi}.
UHECR interactions with gas surrounding the source were considered in~\cite{Muzio:2019leu} but found to be subdominant and therefore we do not include them here. 
Finally, a number of ``nuisance" parameters characterize the Galactic cosmic rays' spectral index, cutoff rigidity and normalization; see~\cite{Unger:2015laa, Muzio:2019leu, Muzio:2021zud} for a more detailed description of the features available in the UFA code.    

For the present work we have extended the UFA framework~\cite{Unger:2015laa,Muzio:2019leu} and code (SourceProp~\cite{UFACode}) 
to include photo-hadronic interactions of heavy UHECR nuclei (mass number $A>56$) with the photon field around the source and during extragalactic propagation.  The required interaction lengths in the source environment are calculated using the cross section and branching ratio tables provided by~\cite{CRPropa:2025trd, Morejon:2021nvz}.  The extragalactic propagation of UHECR nuclei with $A>56$ up to $A=140$ is calculated using SimProp~\cite{Aloisio:2017iyh, diMatteo:2025xod}. We have also extended the code to include two distinct components (jet and bulk outflow) having potentially different temperatures and other properties.
As in previous UFA analyses, the diffuse flux of UHECRs arriving at the Earth makes use of the approximation that each individual source emits the same spectrum of UHECRs. However, in the BNSM scenario this is physically motivated~\cite{fBNS-prl25} whereas it is unjustified for other proposed source classes~\cite{EhlertOikonomouUnger23}. 

The redshift evolution of the BNS merger rate is not yet known observationally. A star formation rate (SFR) evolution would be applicable if the BNS system formed and merged immediately after the stars themselves formed.  However the characteristic time delay between the formation of the BNS system and its merger is likely to be (much) longer, possibly as long as $\tau = 5\;\mathrm{Gyr}$~\cite{Du:2025wto}.  To bracket the possible range, we take the dot-dashed red line in Fig. 1 of~\cite{Du:2025wto}, for which $\tau = 5\;\mathrm{Gyr}$, as the representative of the BNSM evolution. We parameterize the redshift evolution for this ``BNSM" source evolution model as: 
\begin{equation}
\label{eq:BNSM_0}
 \xi^0_\mathrm{BNSM} \propto 
 \begin{cases}
 \exp\left( -\frac{(z-\mu)^2}{(2\sigma)^2}\right)  & \mathrm{for} \quad z \leq 1.5 \\
 \exp\left( -\kappa z \right)  & \mathrm{for} \quad z > 1.5 \ ,\\
\end{cases}
\end{equation}
where $\mu = 0.87$, $\sigma = 0.60$, and $\kappa = 0.88$.
We also report the results using the SFR evolution~\cite{Robertson:2015uda} which represents an upper limit on the high-z production of UHECRs.  We expect that these two evolution models, compared in Fig.~\ref{fig:redshift_evolution}, bracket the range of possibilities.

\subsection*{Data and Procedure}
\label{sec:data_and_procedure}

The 3000 km$^2$ Pierre Auger Observatory~\cite{PAO_NIM_15}, in the Argentine Pampas, is the world's largest UHECR detector; it has been taking data for more than 20 years. The Telescope Array is an important complementary observatory in the northern hemisphere~\cite{TA_FD,TA_SD}, but has considerably lower statistics and less robust composition sensitivity, so we restrict ourselves in this analysis to the Auger data.

The basic elements of the Pierre Auger Observatory are a 3000 km$^2$ surface detector array (the SD) consisting of 1600 water Cherenkov detectors, overlooked by four air-fluorescence telescopes known together as the fluorescence detector (FD).  The FD provides direct, calorimetric energy determination.  In addition, the FD provides sensitivity to the composition of the primary CRs on a statistical basis because the interaction cross section of a heavier primary is larger, so the first interaction occurs on average at smaller column depth, $X$, and the shower of a more massive UHECR develops more quickly and with less event-to-event variance. The mean $\langle X_\mathrm{max} \rangle$ and the event-to-event variance $\sigma(X_\mathrm{max})$ thus encode composition information which is interpreted using particle physics models of air shower development constrained by accelerator data. 
The FD can only operate on dark nights, so hybrid (SD+FD) observation is only available for about 10\% of the events. The energy of SD-only events is determined and their composition constrained by use of the hybrid dataset. The Auger observatory has recently undergone a major upgrade to more accurately determine the composition and to improve our understanding of particle physics at extreme energies; see~\cite{Castellina_AugerII} for more details. For our purposes here, the large dataset of the Auger Phase I detector is key, so we restrict our analysis to that data.  

We extract information about the UHECR accelerator and its environment by finding the parameters of the composition-extended UFA model that give the best fit to the latest Auger spectrum and composition data. We take the spectrum above $10^{18.4}\;\mathrm{eV}$ from~\cite{PierreAuger:2025eun} and at lower energy from ~\cite{PierreAuger:2021hun}; we restrict our fitting to $E>10^{17.5}\;\mathrm{eV}$, since Galactic cosmic rays dominate at lower energies.  For the composition sensitive observables, $\langle X_\mathrm{max} \rangle$ and $\sigma(X_\mathrm{max})$, we use the FD data from~\cite{PierreAuger:2026qbt} below 
energy $10^{18.5}\;\mathrm{eV}$;  at higher energies, we use the high-statistics determination of $\langle X_\mathrm{max} \rangle$ from surface-detector-only events up to $\approx 10^{19.75}\;\mathrm{eV}$, which Auger calibrated from hybrid events using a Deep Neural Network (DNN)~\cite{augerDNNPRD25}. 
As recommended, we make independent overall shifts to  
$\langle X_\mathrm{max}\rangle$ and $\sigma(X_\mathrm{max})$ from the DNN analysis to match the normalization of the corresponding FD data in the overlapping energy range; the needed shift is +5$\;\mathrm{g\;cm^{-2}}$ in both cases. 
The hadronic interaction model EPOS-LHC-R~\cite{Pierog:2023OB} accounts for the data much better than earlier models, largely solving the muon and $\langle X_\mathrm{max} \rangle$ problems.  Therefore we use EPOS-LHC-R with no energy or $\langle X_\mathrm{max} \rangle$ shifts to convert $\langle X_\mathrm{max} \rangle$ and $\sigma(X_\mathrm{max})$ to the mean and variance of the composition, ln$A$, using the parametrization given in~\cite{Evoli:2026bea}.  

We constrain the model parameters by minimizing $\chi^2$, fitting to the spectral and composition data as discussed above.  We follow the procedure of the Particle Data Group~\cite{ParticleDataGroup:2024cfk} to compare the quality of different fits in terms of the effective $\sigma$ significance of the difference in the fits; see the Supplemental Materials for details.  We assess the impact of systematic uncertainties in the energy scale ($\pm14$\%) and depth of shower maximum ($\pm 10$g/cm$^{-2}$), by refitting all of the models considered with the 8 additional combinations of shifted data.

\subsection*{Analysis and Results}
\label{sec:Analysis and Results}

We begin by finding how well an optimized ``generic UFA" UHECR model can fit the latest Auger spectrum and composition data, using the currently-best hadronic interaction model, EPOS-LHC-R~\cite{Pierog:2023OB}.  The result depends on the assumed redshift-evolution of the sources.  A common expectation is that the source evolution mirrors the star formation rate, as for source candidates like long GRBs. We call this the ``no-jet SFR" model; its $\chi^2$/ndf is 5.52. The fit improves if the ``BNSM" source evolution discussed above is used, instead of the SFR evolution.  
While a low-z-peaked source evolution is not obviously well-motivated for alternate UHECR source models, to be conservative in judging the relative quality of fit in the BNS merger scenario we use the ``no-jet BNS" model as the baseline for representing non-BNSM models; its $\chi^2$/ndf is 4.49.  The fit parameters for both the ``no-jet SFR" and ``no-jet BNS" models are given in Table~\ref{tab:1}.

\begin{figure}
\centering
\includegraphics[width=0.48\textwidth]{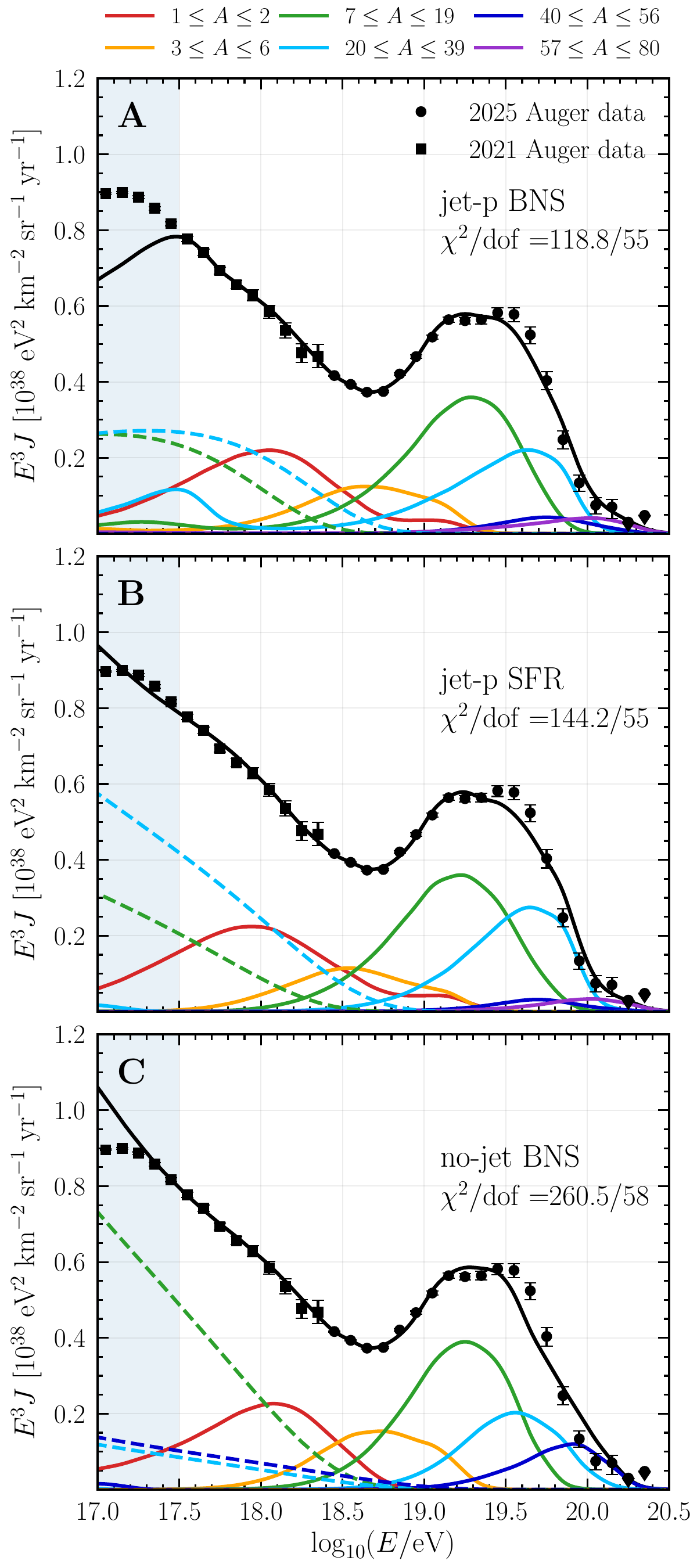}
\includegraphics[width=0.48\textwidth]{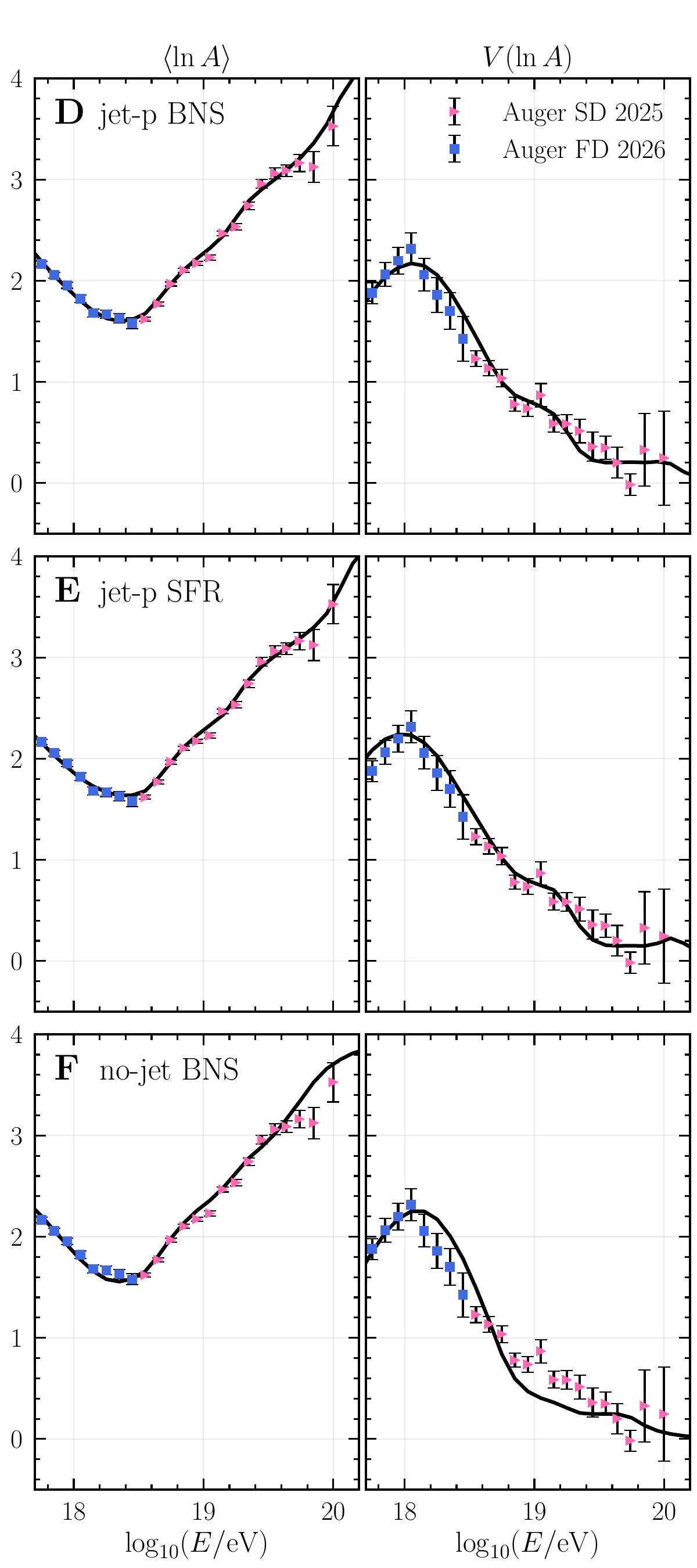}
\caption{\textbf{Testing the BNSM model with Auger data.} The top panels show the BNSM predictions (``jet-p BNS" in Table~\ref{tab:1}) for the spectrum (left) and composition observables (right) compared to the Auger data.  Comparing to the middle panels, which assume SFR source evolution, shows that the BNSM predictions for UHECR observables are fairly insensitive to the redshift evolution of the BNS merger rate.  The bottom panels show the best fit to the data in the corresponding non-BNSM scenario without the light jet; it is worse at 7.6$\sigma$ significance.}
\label{fig:Figure1}
\end{figure}

Unlike other proposed UHECR source types, the BNSM scenario predicts the existence of two distinct components:  the UHECRs accelerated in the jets and those accelerated in the magnetized turbulent outflow at larger angles~\cite{fBNSApJL25}. Our goal here is to ascertain whether or not the data shows evidence for both these components. It was argued in~\cite{fBNSApJL25} that most UHECRs are accelerated in the bulk outflow, where nucleosynthesis of heavier nuclei occurs.  The  cutoff rigidities in the outflow were estimated in~\cite{fBNSApJL25} by balancing the acceleration and synchrotron loss timescales, using the initial conditions on the magnetic field from the super-high-resolution GRMHD-with-neutrinos simulation of a BNSM~\cite{Kiuchi24}.  The resultant estimated cutoff rigidities for $A=1, 14, 28, 56 $ and the first r-process peak with $A=80$, are respectively 6.2, 7.3, 6.4, 6.0 and 5.9 EV.  These values are close enough that it is adequate to model the nuclei accelerated in the bulk outflow as having a common rigidity cutoff.

Particle-in-cell simulations of acceleration in magnetized turbulence have shown that the spectrum is a power-law with index near -2 and a sech$[-(R/R_{\rm max})^2]$ cutoff~\cite{cfm24}, so we include five UHECR mass groups with normalizations $f_A$, a sech rigidity cutoff $R_{\rm max}$, and spectral index $\gamma$.
As previously noted, the UFA analysis framework  takes into account their interactions with the ambient photon field.  These interactions have a large impact on the spectrum and composition escaping the source over a significant portion of the energy range, as discussed below and exemplified by the plots in Fig.~\ref{fig:Figure1-appandix}. We take the photon field to be blackbody and have checked that modifications in the form of an absorption factor (see Appendix A of~\cite{Unger:2015laa}) or a broken power law do not significantly impact the analysis.

Here we assume that no single source makes a prominent contribution to the observed UHECRs.  However since the jet opening angle is small, $10-12^\circ$, future analyses combining directional anisotropies as well as spectrum and composition may need to take into account that -- for individual sources -- the relative contribution of the jet will vary from source to source depending on viewing angle and distance.

\subsubsection*{High energy light component of UHECRs}
\label{sec:He_and_p}

In addition to the bulk UHECRs, we allow for a separate higher energy light component, which is a signature feature of the BNSM scenario~\cite{fBNSApJL25}. We call it the ``jet" component for brevity, but the present analysis is not sensitive to the physical origin of the higher-energy light component. Reference~\cite{Muzio:2019leu} found that a subdominant high energy light component, of unspecified origin, could improve the fit substantially, depending on the hadronic interaction model.  However the hadronic interaction models of those days were in significant tension with multiple features of the air shower data, so a firm conclusion could not be drawn.  The present spectrum and composition data~\cite{PierreAuger:2021hun, PierreAuger:2025eun, augerDNNPRD25, PierreAuger:2026qbt} are far more constraining, particularly for composition at the highest energies where the Auger DNN analysis~\cite{augerDNNPRL25,augerDNNPRD25} increased the statistics by an order of magnitude.  Furthermore, EPOS-LHC-R~\cite{Pierog:2023OB} is much better at fitting accelerator and air shower data than the hadronic interaction models available in 2019, reducing the observational uncertainty on composition. 

As a baseline model for the BNSM scenario, we take the jet component to be pure proton, as is expected if collapse to a black hole is immediate without a transient hypermassive neutron star~\cite{Perego+LightElems22}.  In our baseline model possible interactions of the jet component are ignored, a good first approximation since the proton energy loss length is large compared to that of nuclei. Results including processing in the photon field of the jets and considering helium  accelerated in the jets will be reported elsewhere.  In a preliminary exploration of such processing, we have found that it does not change the conclusions  presented here. 

The spectrum of the high-energy jet component, $\phi^\mathrm{jet}$, is parameterized as a power law with spectral index $\gamma^\mathrm{jet}$ and cutoff energy $E^\mathrm{jet}$.  PIC simulations for proton acceleration in DSA have not yet been done, so we consider both a sech cutoff as found in PIC simulations for magnetized turbulence~\cite{cfm24} and an exponential cutoff as predicted analytically for DSA~\cite{ProtheroeStanev99}.  Their fit qualities are similar, but sech gives a better fit and facilitates comparison of the cutoff values for the two components, so unless otherwise stated we show results for a sech cutoff and drop the ``sech" in the model name unless needed to contrast with a fit with ``exp" cutoff.  
We quantify the contribution of the jet (whose spectral index may be different than that of the bulk outflow) by the fraction of energy it carries, relative to all UHECRs, above the reference energy $E_\mathrm{ref} = 10\;\mathrm{EeV}$: 
\begin{equation}
\label{eq:fraction_of_secondary_UHECR}
f^\mathrm{jet} = \frac{\int_{E_\mathrm{ref}}^{\infty} dE \; E \;\phi^\mathrm{jet}}{\int_{E_\mathrm{ref}}^{\infty} dE \; E \; (\phi^\mathrm{jet} + \phi)}  \ .
\end{equation}

Figure~\ref{fig:Figure1} shows the spectrum and composition fits at Earth with and without a high-energy jet proton component (top and bottom rows, respectively), using BNSM source evolution.  For comparison, in the middle row we show the results with a jet proton component having sech cutoff, using SFR source evolution.  
The impact of the interactions of the bulk component in the photon field is shown in Fig.~\ref{fig:Figure1-appandix}. 
The parameters for the best fits with these three models, and also for an exp cutoff with BNSM evolution, are given in Table~\ref{tab:1}.   
The fit with a $+1 \sigma$ systematic shift in the absolute energy scale is also given in Table~\ref{tab:1} for reference.
Tables~\ref{tab:jet-p-BNS-shifts1}-~\ref{tab:no-jet-SFR-shifts2} report the fits for the models ``jet-p BNS" and ``no-jet SFR", for all combinations of systematic shifts.   

Table~\ref{tab:1} shows that the basic BNSM models, with BNS source evolution and either sech  (or exponential) cutoff, fit significantly better than the best-fitting no-jet model, by $7.6 (7.2) \sigma$.  The nine combinations of $0, \pm 1 \sigma$ shifts in energy and $X_{\rm max}$ scales, generally produce little change or a worse fit, except that the fit becomes even better for the ``jet-p BNS shift" model, with a $1 \sigma$ upward shift in energy scale. This case is interesting since a small upward shift in the energy calibration would improve consistency of the Auger and TA spectra~\cite{SpectrumWG19}, moreover the fitted jet spectral index, $\gamma^\mathrm{jet} = -1.8$, is better aligned with conventional expectations than the $\approx 0$ value found without the shift. The steeper spectral slope is, however, not an argument for the shifted energy scale, since simulations of DSA have shown that hard spectral indices arise in certain geometries of magnetic field relative to the shock. 

With or without a high-energy proton component at the source, the contribution of primary protons in the bulk outflow is very small or negligible, as expected according to BNS nucleosynthesis simulations such as~\cite{Perego+LightElems22,radice+BNSnucleosyn22}~\cite{fBNSApJL25}. 

We focused here on the case that the jets contain only protons, as predicted for prompt collapse of the hypermassive neutron star transient produced in the binary neutron star merger.  An interesting question is whether UHECR data can determine whether the jet may also -- or, conceivably, preferentially -- accelerate helium, which may be produced when the neutron star equation of state is stiff enough to delay collapse. Because helium is  readily photo-dissociated, both in the jet and in extragalactic propagation, discriminating a primary He population appears difficult. We have also investigated whether interactions of jet protons with the radiation field in the jet can leave a sufficiently distinctive imprint on the UHE protons arriving at Earth to usefully probe conditions in the jet, but that also appears difficult. Results of both studies will be presented elsewhere.

\subsubsection*{Possible r-process UHECRs}
\label{sec:Fraction of r-process nuclei in the UHECR fulx}

\begin{figure}[t]
\centering
\includegraphics[width=0.96\textwidth]{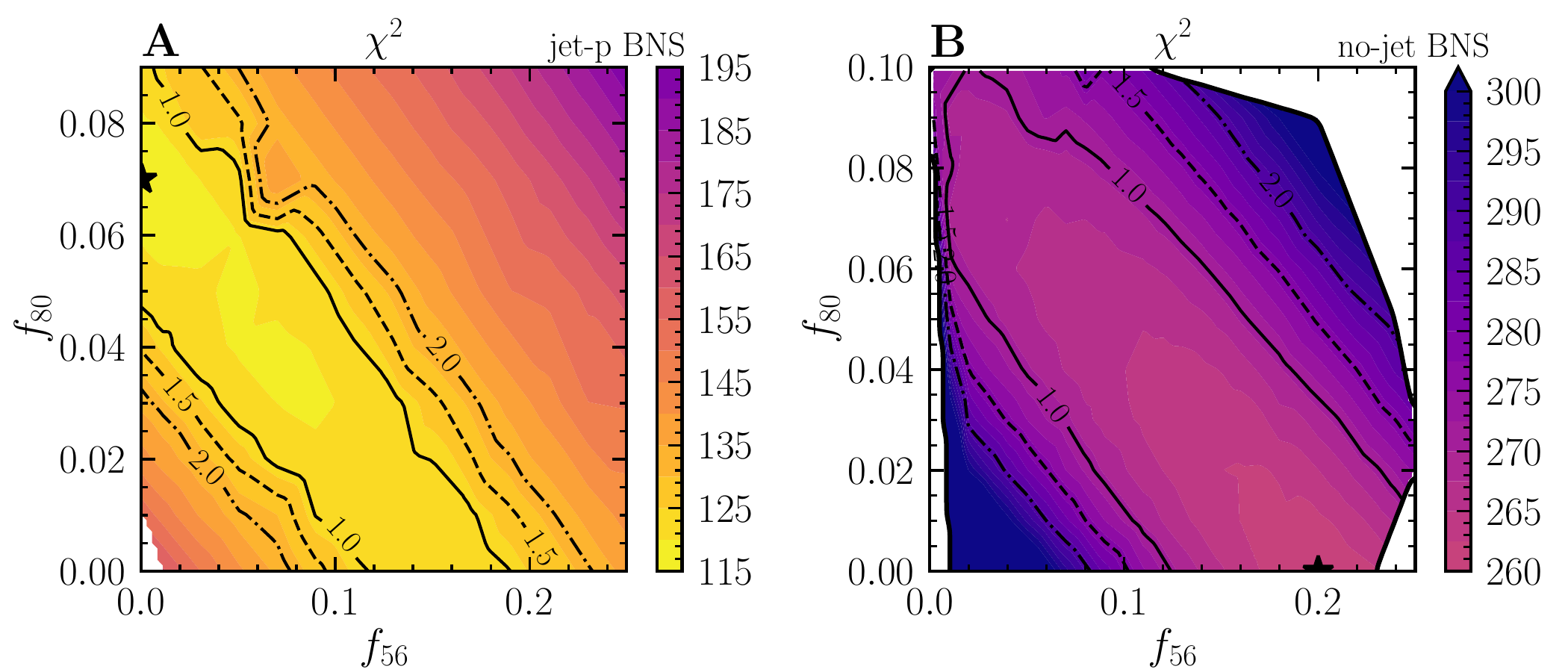}
\caption{\textbf{Fit quality ($\chi^2$) as a function of the heavy-fractions.} 
Best-fit model ``jet-p BNS" (panel~\textbf{A}) and ``no-jet BNS" model (panel \textbf{B}).  
}
\label{fig:2}
\end{figure}

Another distinctive feature of the BNSM model for UHECR production is the possible presence of nuclei heavier than iron~\cite{fBNS-prl25} and we have allowed for that possibility in our study~\cite{Zhang-note}.
In fact, all of the jet models prefer a non-zero component of $A>56$ at the source (see Table~\ref{tab:1}).  Indeed, for the favored BNSM-motivated ``sech jet-p BNS" case, there is no iron and the fraction of heavy nuclei is $f_{80} = 7\%$.  To see how well-constrained the apportionment between iron and r-process nuclei is, we performed the fits for a grid of $f_{56}$ and $f_{80}$ values. Fig.~\ref{fig:2} shows the resulting 2D plot of $\chi^2$ for the BNSM exemplar ``jet-p BNS" model and the non-BNSM ``no-jet BNS"model. 
These plots show that with a jet, the data favor a significant component of r-process origin, but does not require~it.  

Figure~\ref{fig:2D-appandix} in the Supplemental Material explores how the favored injection index and temperature depend on $f_{56}$ versus $f_{80}$. The index is around -2.1, independent of how the heavy (A=56+A=80) the bulk composition is partitioned. The inferred temperature of the bulk outflow photon field is similarly insensitive to the weighting within the heavy component.  The temperature values, around 600K, are lower than the reported temperatures of the kilonova of GW170817, but more detailed study is required to know if this is a tension or is due to the temperature in the UHECR-escape region not being probed by the kilonova observations. We also note that the UFA framework simplifies the temperature of the photon field as being uniform, whereas in the real system it is undoubtedly spatially-varying.   

\subsubsection*{Neutrino spectrum and redshift evolution of BNSMs}
\label{sec:Neutrinos}

\begin{figure}[t]
\centering
\includegraphics[width=0.9\textwidth]{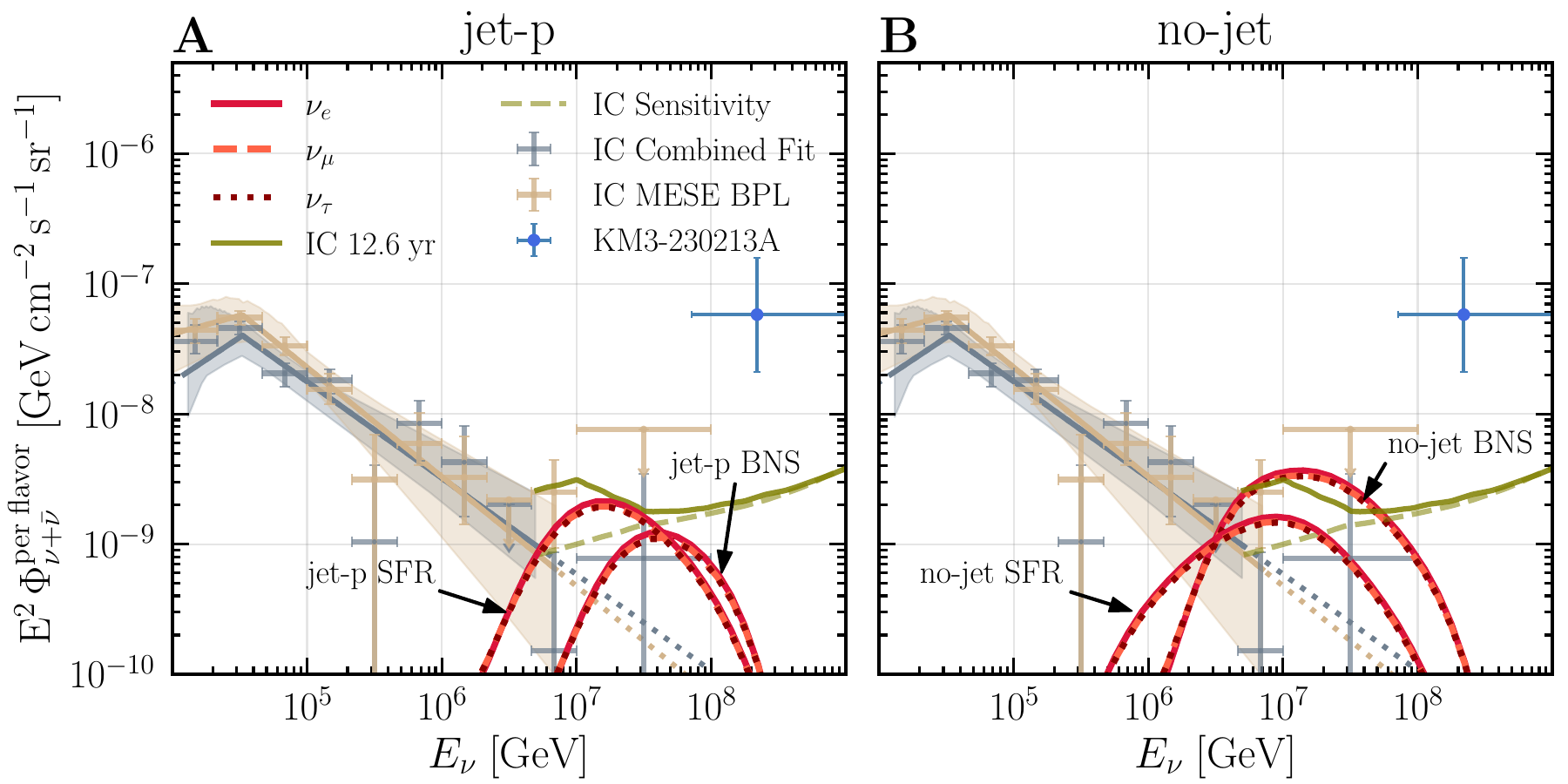}
\caption{\textbf{Diffuse neutrino fluxes originating from UHECR interactions compared to data} for the main fit scenarios considered here: jet-p BNS (panel \textbf{A}), jet-p SFR (panel \textbf{A}), no-jet BNS (panel~\textbf{B}), and no-jet SFR (panel~\textbf{B}). We also show the IceCube data points from their Combined Fit and MESE analysis with a broken power law~\cite{IceCube:2025tgp}, their 90\% CL upper limits from 12.6~yr of exposure, and their sensitivity~\cite{IceCubeCollaborationSS:2025jbi}. The high-energy KM3NeT event is also shown~\cite{KM3NeT:2025npi}.
}
\label{fig:neutrinos}
\end{figure}

Due to the GZK energy loss ``horizon"~\cite{Zatsepin:1966jv, Greisen:1966jv}, the UHECRs observed by Auger above the ankle were mostly produced within the last few hundred megayears.  However UHECRs' progeny neutrinos -- produced as a result of the photo-disintegration interactions discussed above -- only lose energy through cosmological redshifting, so their spectrum integrates the contribution of UHECR production over its full redshift range.  Hence the UHE neutrino spectrum ($E\gtrsim$few PeV) is directly sensitive to the redshift evolution of BNS mergers. Figure~\ref{fig:neutrinos} shows the spectrum of neutrinos predicted for the three different best fit scenarios from Fig~\ref{fig:Figure1} and the no-jet SFR model, compared with IceCube observations and the KM3NeT event.  For the jet scenarios (panel A), one sees clearly how the spectrum of neutrinos produced by UHECRs, integrated over cosmic time, has a higher peak energy and overall smaller amplitude for BNS source evolution because the BNS merger rate peaks later (at lower redshift) than SFR (Fig.~\ref{fig:redshift_evolution}). 
The difference between neutrinos in the no-jet scenarios relative to the corresponding jet scenario, follows from the need for a larger number of photo-disintegration interactions (larger $\langle N_{\rm int} \rangle$) and higher temperature to approximate the UHECR data without the jet contribution.  These respectively increase the number of neutrinos per cosmic ray and add lower energy neutrinos.  Hence the spectrum of UHE neutrinos, when measured, will be a good probe of the cosmic evolution of the neutron star merger rate, and further test the BNSM origin of UHECRs. (See~\cite{Guo+25} for a different picture of UHE neutrino production in BNSMs.)

\subsection*{Summary and Conclusions} 
\label{sec:Conclusions}

We have simulated the spectrum and composition at Earth, of UHECRs produced in binary neutron star mergers, and compared the predictions to the latest observations from the Pierre Auger Observatory. There are two distinct populations of UHECRs: a light component originating in the jets, and a heavier component accelerated in the magnetized bulk outflow with masses potentially reaching into the r-process regime~\cite{fBNSApJL25}.  Both populations are modified by their interactions with photons prior to reaching Earth, which we account for using the UFA code written by M.~Unger~\cite{Unger:2015laa}, extended by us to include CRs with masses up to $A=140$.  Photo-disintegration of the heavy population generates a secondary population of protons whose maximum energy is lower than for the jet protons.

The data strongly supports, at $\geq 7.2 \sigma$, the presence of the predicted two populations of protons. The fraction of energy carried by jet protons relative to all CRs above 10 EeV is $\approx$5-10\%.  One can see by eye in comparing panels {\bf A} vs. {\bf C} and {\bf D} vs. {\bf F} of Fig.~\ref{fig:Figure1}, how the presence of the jet protons naturally contributes subtle structures in the data which were previously unexplained.  The cutoff rigidity of the bulk outflow is found to be 7.2 EV and the cutoff energy of the jet protonic component is $\approx 12$ EeV. 
These values are in remarkable agreement with the simple, no-free-parameter estimates of 6.6 EV for the outflow (averaging the values in~\cite{fBNSApJL25} using the fractions $f_A$ in Table~\ref{tab:1}) and $\geq$11.5 EeV for the jets~\cite{fBNSApJL25}. 

We investigated whether the UHECR data provides evidence for or against the presence of an r-process component at the source.  Finding r-process nuclei in UHECR data would be smoking-gun evidence for the BNS origin of UHECRs, but predicting the expected level of r-process nuclei accelerated to ultrahigh energy requires detailed modeling of the fragmentation of accelerated nuclei within the expanding ejecta, which has not yet been tackled~\cite{fBNSApJL25}. Our best fit modeling finds an intriguing 7\% of $A=80$ at the source, but the fit imposing $A\leq 56$ is 
within $1\sigma$ of the best fit.
Improvements in interpreting the UHECR data coming from the enhanced composition information on individual events with Auger Phase II~\cite{Castellina_AugerII}, and from the resulting improved hadronic interaction modeling, may enable a sharper conclusion to be drawn within the next few years.  
The photo-disintegration interactions modeled here, which shape the UHECR spectrum and composition, produce neutrinos.  Those neutrinos reach Earth unattenuated, but with redshifted energies.  Thus the spectrum of UHE neutrinos, $E_\nu \gtrsim$few PeV, when measured, will constrain the redshift dependence of the binary neutron star merger rate. 

In summary, the origin of UHECRs in neutron star mergers is corroborated at $>\!\!7\sigma$ by the spectrum and composition data of the Pierre Auger Observatory. The data slightly favors ($1 \sigma$) some UHECRs originating as r-process nuclei.

\begin{table}
\centering
\caption{\textbf{The best fit parameters for the source models considered.} The first rows give $\chi^2/\mathrm{ndf}$ and the number of $\sigma$  preference relative to the no-jet BNS model.  } 
\label{tab:1} 
\begin{tabular}{l c c c c c c }
\\
\hline
\hline
Parameter  & jet-p BNS sech & jet-p BNS exp & jet-p BNS shift & jet-p SFR & no-jet BNS & no-jet SFR \\
\hline
\hline
$\chi^2/\mathrm{ndf}$ & $118.8/55 =$ & $124.3/55 =$ & $99.9/55 = $ & $144.2/55 = $ & $260.5/58 =$ & $320.2/58 = $ \\
{} & $2.16$ & $2.18$ & $1.82$ & $2.62$ & $4.49$ & $5.52$ \\
$N_\sigma$ & $7.6$ & $7.2$ & $8.9$ & $6.1$ & $-$ & $-3.7$ \\
$\lg \langle N_\mathrm{int} \rangle$ & $2.55$ & $2.58$ & $2.53$ & $2.99$ & $3.14$ & $2.59$ \\
$\gamma_\mathrm{inj}$ & $-2.11$ & $-2.12$ & $-2.13$ & $-1.90$ & $-2.14$ & $-1.25$ \\
$\lg (R_\mathrm{max}/\mathrm{V})$ & $18.86$ & $18.85$ & $18.92$ & $18.88$ & $18.85$ & $18.81$ \\
$\lg (R_\mathrm{diff}/\mathrm{V})$ & $18.01$ & $17.93$ & $18.04$ & $15.58$ & $15.89$ & $16.63$ \\
$r_\mathrm{size}$ & $7.0$ & $6.7$ & $7.4$ & $\gtrsim100$ & $\gtrsim100$ & $\gtrsim100$ \\
$T$ [K] & $563$ & $568$ & $578$ & $756$ & $1235$ & $6044$ \\
$\lg (E^\mathrm{jet}/\mathrm{eV})$  & $19.08$ & $18.72$ & $19.34$ & $19.12$ & $-$ & $-$ \\
$\gamma^\mathrm{jet}$ & $\approx 0$ &  $\approx 0$ & $-1.81$ &  $\approx 0$ & $-$ & $-$ \\
$\lg(f^\mathrm{jet})$ & $-1.31$ & $-1.29$ & $-1.15$ & $-1.19$ & $-$ & $-$ \\
$f_{1}$ &  $\approx 0$ &  $\approx 0$&  $\approx 0$ &  $\approx 0$ & $0.02$ & $0.33$ \\
$f_{14}$ & $0.27$ & $0.27$ & $0.07$ & $0.21$ & $0.65$ & $0.52$ \\
$f_{28}$ & $0.66$ & $0.62$ & $0.82$ & $0.75$ & $0.13$ & $0.13$ \\
$f_{56}$ &  $\approx 0$ & $0.07$ & $4\times10^{-3}$ &  $\approx 0$ & $0.20$ & $\approx 0$ \\
$f_{80}$ & $0.07$ & $0.04$ & $0.10$ & $0.04$ &  $\approx 0$ & $0.01$ \\
$f_\mathrm{Gal}$ & $0.63$ & $0.63$ & $0.67$ & $0.77$ & $0.83$ & $0.85$ \\
$\gamma_\mathrm{Gal}$ & $-2.88$ & $-2.87$ & $-2.96$ & $-3.19$ & $-3.23$ & $-3.26$ \\
$\lg( E^\mathrm{Gal}_\mathrm{max}/\mathrm{eV})$ & $18.49$ & $18.51$ & $18.52$ & $18.60$ & $18.75$ & $18.67$ \\
$f^\mathrm{Gal}_{14}$ & $0.51$ & $0.54$ & $0.23$ & $0.36$ & $0.75$ & $0.57$ \\
$f^\mathrm{Gal}_{28}$ & $0.49$ & $0.46$ & $0.77$ & $0.64$ & $0.12$ & $0.39$ \\
$f^\mathrm{Gal}_{56}$ &  $\approx 0$& $\approx 0$ &  $\approx 0$ &  $\approx 0$ & $0.13$ & $0.04$ \\

\hline
\end{tabular}
\end{table}

\clearpage 

\bibliography{Refs} 
\bibliographystyle{sciencemag}

\section*{Acknowledgments}
We thank M. Unger for making the UFA code publicly available, M. Muzio for generous help in using it, and both of them for valuable comments on the manuscript. We are also grateful to A.~di~Matteo for sharing the SimProp code handling heavy nuclei with us and L.~Morejon for sharing the cross section and branching ratio tables for heavy nuclei with us. In addition, A.M.S thanks R.~A.~Batista for answering questions about CRPropa and G.R.F. thanks F.~Oikonomou, D.~Schmidt, J~Vicha, A.~Watson and members of the Pierre Auger Collaboration for input and discussions.
\paragraph*{Funding:}
The work of A.M.S. is supported by the Neutrino Theory Network Program Grant DE-AC02-07CHI11359 and that of G.R.F. is supported by National Science Foundation grant PHY-2413153.
\paragraph*{Data, code and materials availability:}
The Pierre Auger Observatory data used in this work is available in the respective references: spectral data - Table~10~Ref.~\cite{PierreAuger:2021hun} and Table~V~Ref.~\cite{PierreAuger:2025eun} and the composition data -  Table~IV~Ref.~\cite{augerDNNPRD25} and Table~V~Ref.~\cite{PierreAuger:2026qbt}. 
Code by Unger and Muzio implementing the UFA analysis and subsequent improvements has been made publicly available at github~\cite{UFACode}.
\paragraph*{Competing interests:}
There are no competing interests to declare.

\subsection*{Supplementary materials}
Statistical Analysis\\
Figures~\ref{fig:Figure1-appandix} to~\ref{fig:heavies}\\
Tables~\ref{tab:jet-p-BNS-shifts1} to~\ref{tab:no-jet-SFR-shifts2}


\newpage

\renewcommand{\thefigure}{S\arabic{figure}}
\renewcommand{\thetable}{S\arabic{table}}
\renewcommand{\theequation}{S\arabic{equation}}
\renewcommand{\thepage}{S\arabic{page}}
\setcounter{figure}{0}
\setcounter{table}{0}
\setcounter{equation}{0}
\setcounter{page}{1}

\begin{center}
\section*{Supplementary Materials for\\ \scititle}
Anna~M.~Suliga$^{1\ast\dagger}$,
Glennys~R.~Farrar$^{1\ast\dagger}$\\
\small$^\ast$Corresponding authors. Emails: a.suliga@nyu.edu and gf25@nyu.edu\\
\small$^\dagger$These authors contributed equally to this work.
\end{center}

\subsubsection*{This PDF file includes:}
\begin{figure}[t]
\centering
\includegraphics[width=0.94\textwidth]{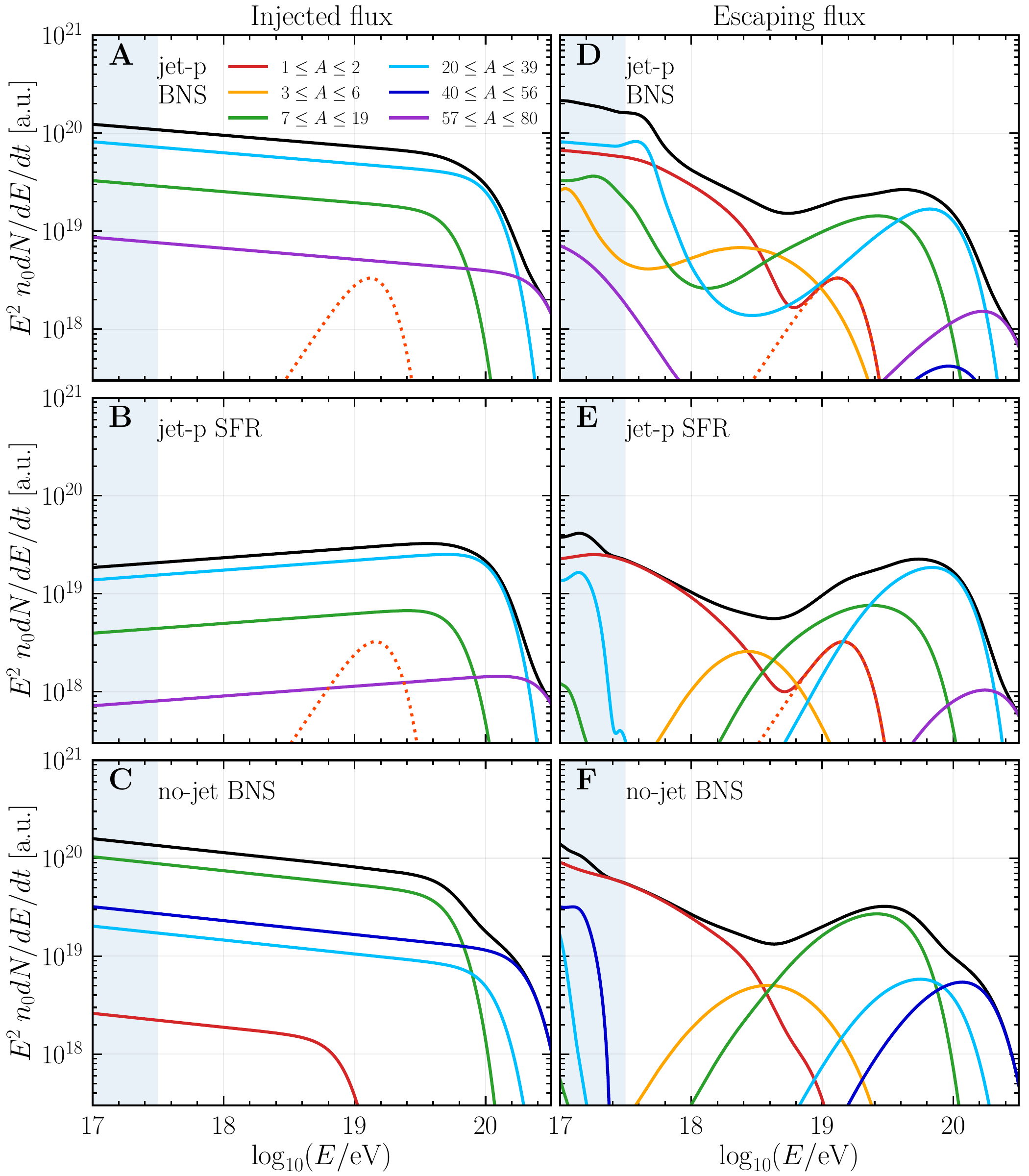}
\caption{\textbf{The impact of photon interactions in the outflow.} 
\emph{Left column:} The injected spectrum of cosmic rays of different mass groups accelerated in the outflow (solid lines) and in the jet proton component (dotted line), for illustrative models. (\textbf{A}) Best-fitting model (jet-p BNS), (\textbf{B}) jet-p SFR and (\textbf{C}) no-jet BNS. \emph{Right column:} The respective spectra taking into account the interaction of UHECRs with the ambient photon field in the bulk outflow (panels \textbf{D}-\textbf{F}).}
\label{fig:Figure1-appandix}
\end{figure}

\begin{figure}
\centering
\includegraphics[width=0.5\textwidth]{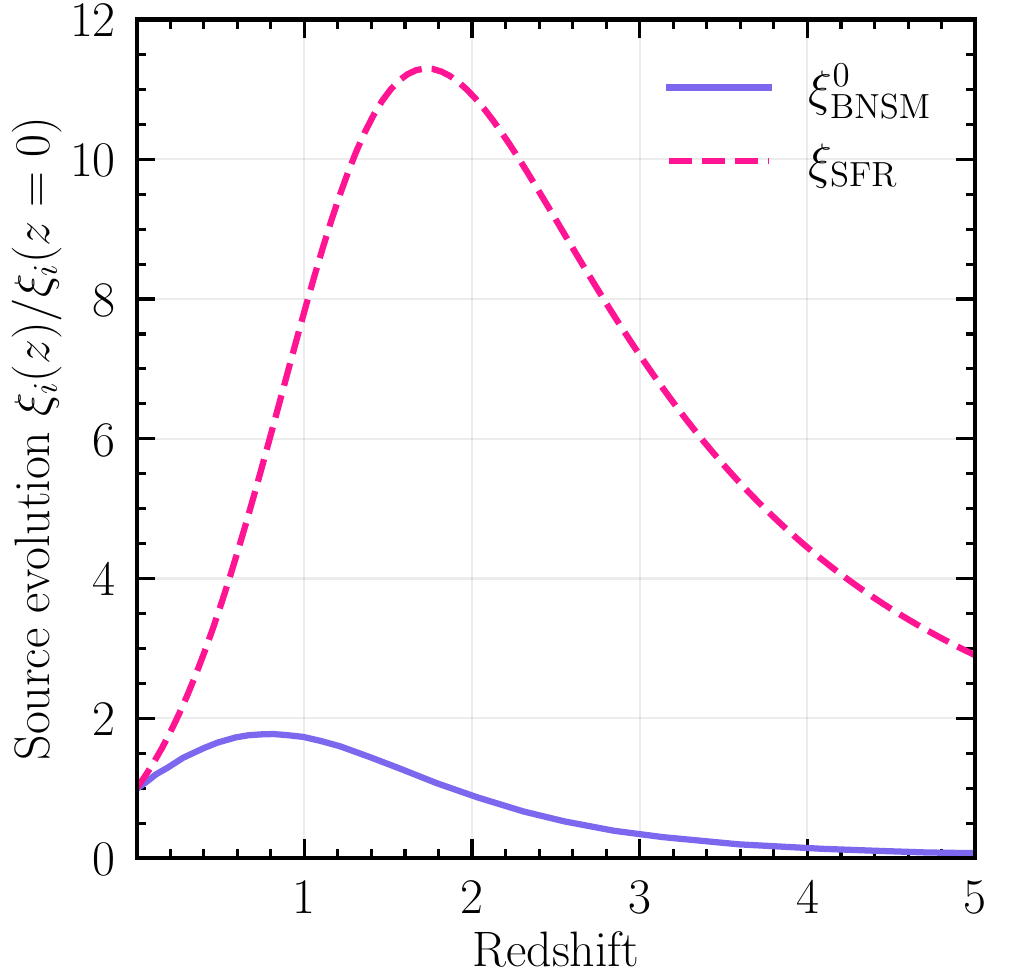}
\caption{
\textbf{The BNSM and SFR source redshift evolutions, normalized to 1 at z=0.} From Refs.~\cite{Du:2025wto} and~\cite{Robertson:2015uda} respectively. 
}
\label{fig:redshift_evolution}
\end{figure}

\begin{figure}[t]
\centering
\includegraphics[width=0.94\textwidth]{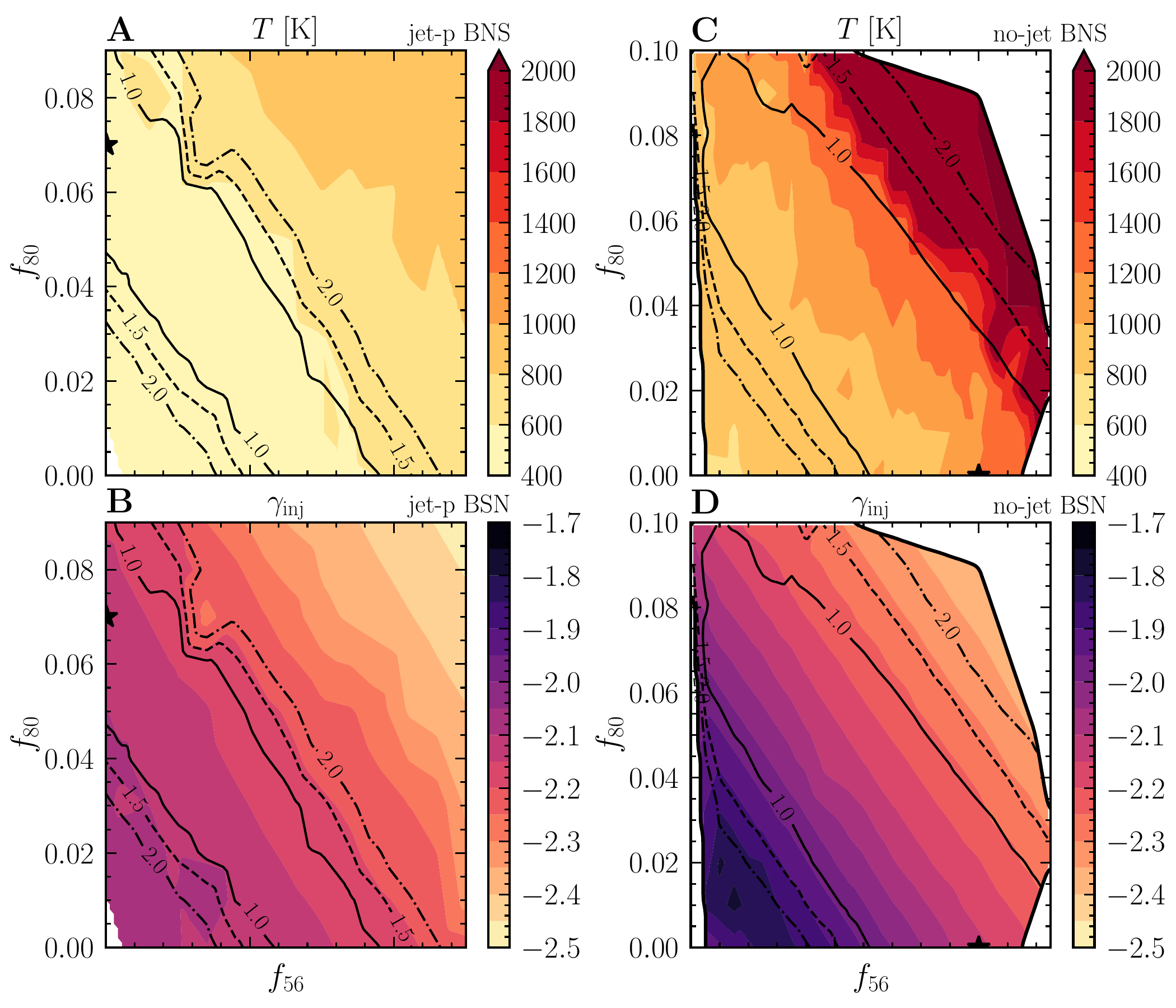}
\caption{\textbf{The temperature and spectral index of the bulk outflow, as a function of the heavy-fractions.} 
\emph{Left column:} Best-fit model (``jet-p BNS") fits:~\textbf{(A)} temperature of bulk outflow photon field, and~\textbf{(B)} power law index of the accelerator, for specified fractions of $A$=56 and $A$=80 at the source. \emph{Right column:} Same, but for the no-jet BNS model (panels \textbf{C}-\textbf{D}). The stars mark the best-fit point for each column. These and other properties of the outflow are relatively insensitive to the presence or absence of an r-process component, for a given $f_{56}+f_{80}$. 
}
\label{fig:2D-appandix}
\end{figure}

\begin{figure}
\centering
\includegraphics[width=0.4\textwidth]{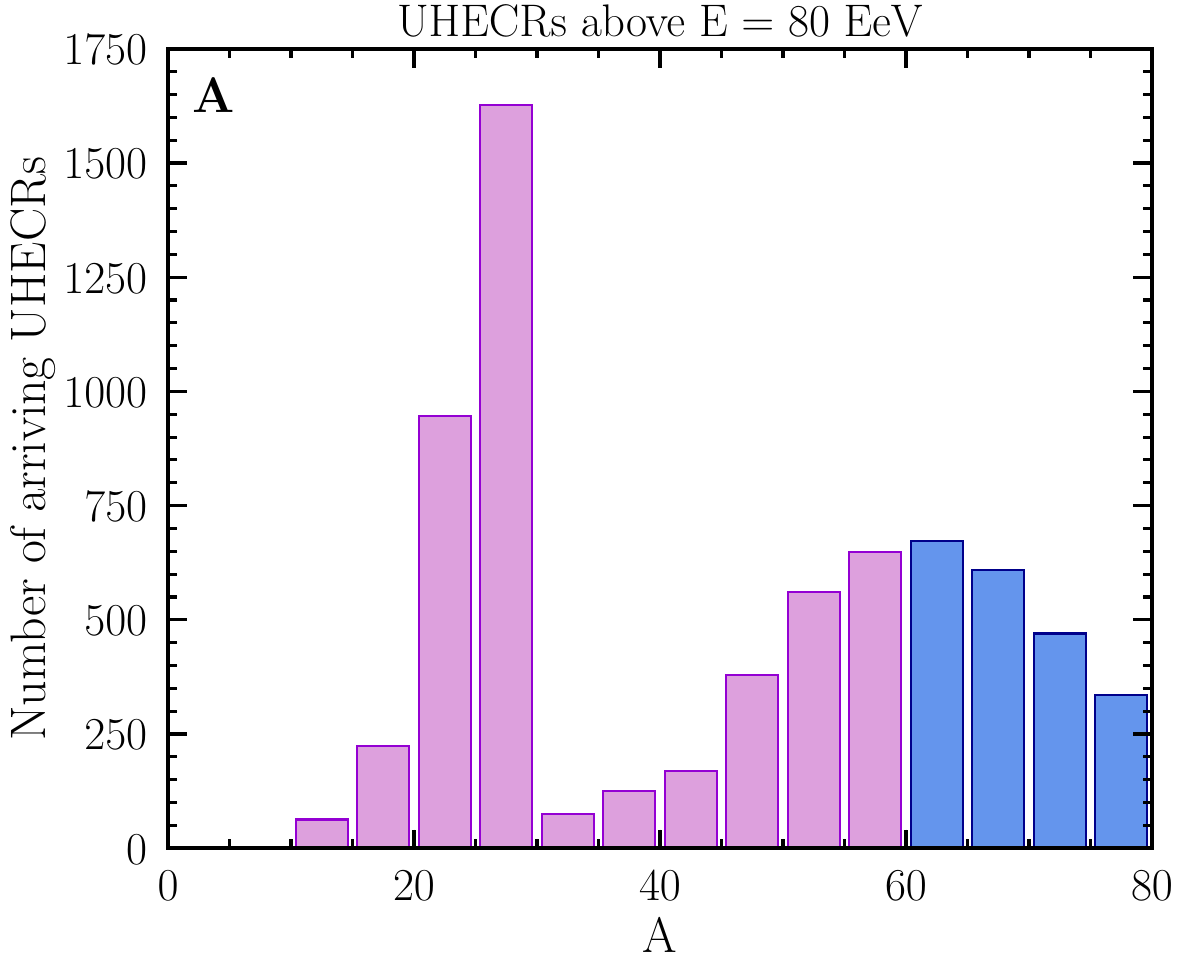}
\includegraphics[width=0.4\textwidth]{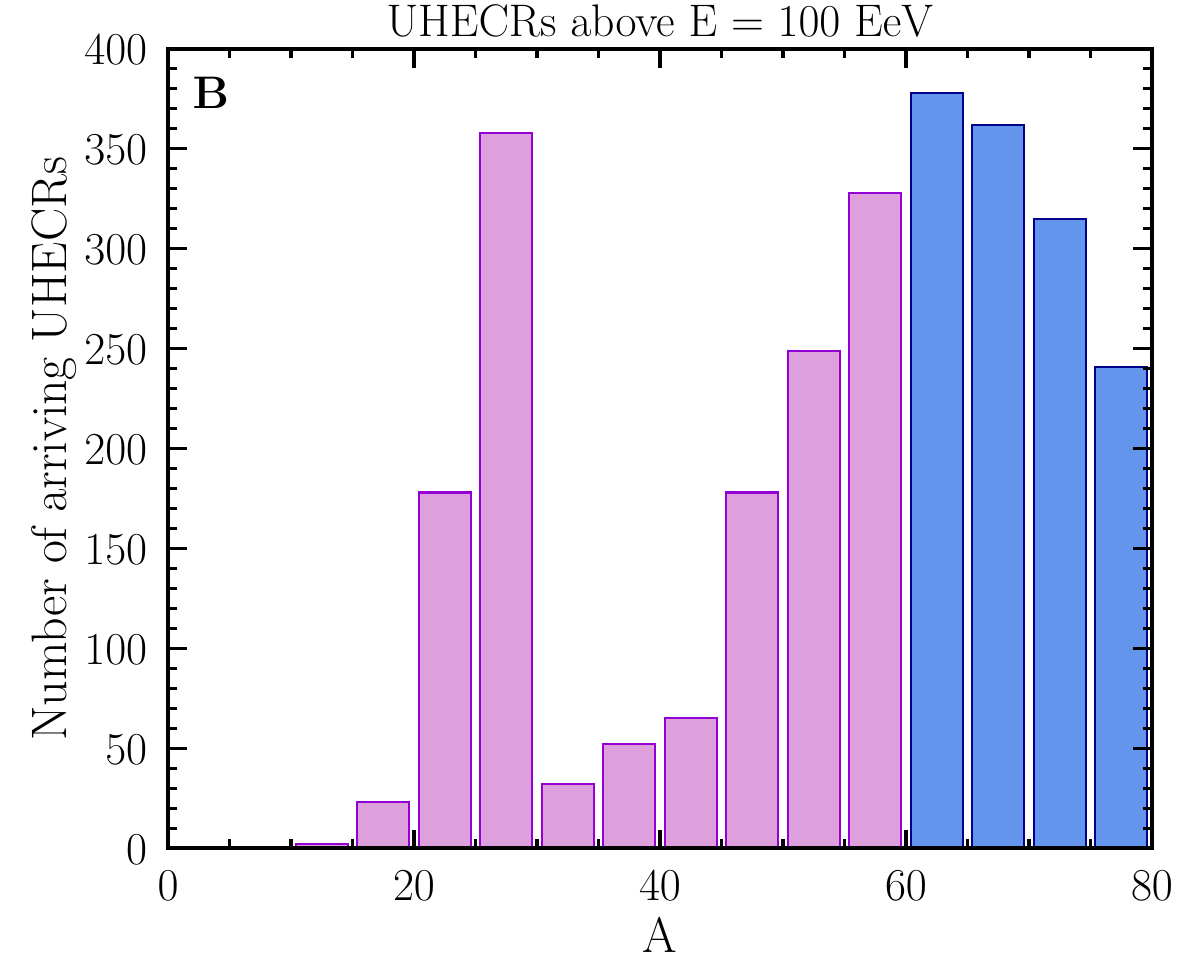}
\includegraphics[width=0.4\textwidth]{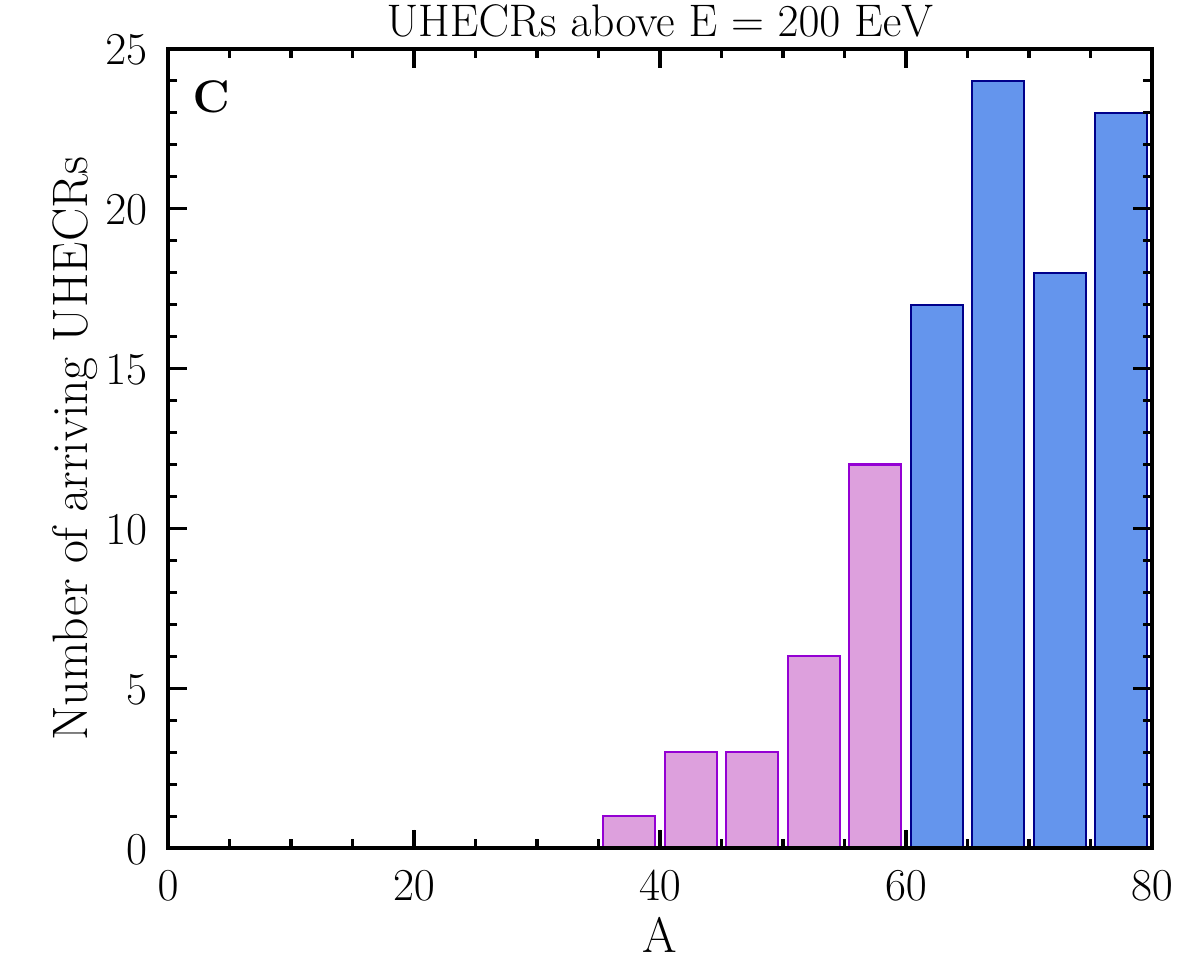}
\caption{
\textbf{The number of events in bins of A with 100x present data, in the baseline ``jet-p BNS" model for various energy thresholds.}  We predict the current Auger dataset contains $\approx\!21$ events with $A>60$ above 80 EeV (panel \textbf{A}), $\approx\!13$ events with $A>60$ above 100 EeV (\textbf{B}), and $\lesssim \!1$ event above 200 EeV (\textbf{C}).  With improved hadronic interaction modeling and better mass determination, it may be possible for this signal to be detected in Auger Phase II. 
}
\label{fig:heavies}
\end{figure}

Statistical Analysis\\
Figures~\ref{fig:Figure1-appandix} to~\ref{fig:heavies}\\
Tables~\ref{tab:jet-p-BNS-shifts1} to~\ref{tab:no-jet-SFR-shifts2}

\section*{Statistical Analysis}
\label{sec:statistical}
We constrain our models by minimizing $\chi^2$ following~\cite{Unger:2015laa, Muzio:2019leu, Muzio:2021zud, Muzio:2022bak}, fitting the spectral and composition data with 
\begin{equation}
\label{eq:chi-test}
\chi^2 = \sum_i^N \frac{\left(J_i - \bar{J_i} \right)^2}{\sigma_{\bar{J_i}}^2} + \sum_j^M \frac{\left(\langle \ln{A}\rangle_j - \overline{\langle \ln{A}\rangle}_j \right)^2}{\sigma_{\overline{\langle \ln{A}\rangle}_j}^2}  + \sum_j^M \frac{\left(\mathrm{Var}(\ln{A})_j - \overline{\mathrm{Var}(\ln{A})}_j \right)^2}{\sigma_{\overline{\mathrm{Var}(\ln{A})}_j}^2} \ ,
\end{equation}
where $N$ is the number of spectral bins, $M$ is the number of composition bins, $\bar{J}$ is the flux, $\overline{\langle \ln{A}\rangle}$ is the mean logarithmic mass, $\overline{\mathrm{Var}(\ln{A})}$ is the variance of the mean logarithmic mass (quantities with overline are the measured ones, those without are the model predictions), and the $\sigma^2_{\bar{\mathcal{Q}}_j}$ are the statistical errors on the observables $\overline{\mathcal{Q}} = \{\bar{J_i}, \overline{\langle \ln{A}\rangle}, \overline{\mathrm{Var}(\ln{A})}\}$.

Our final $\chi^2$ also includes an additional contribution from the spectral energy bins above the highest energy data point in Pierre Auger Observatory spectral dataset~\cite{PierreAuger:2025eun}. We model that contribution following~\cite{Baker:1983tu}, therefore, $\chi^2 = \chi^2 + \sum_i 2n_i$, where $n_i$ is the number of expected events in the Pierre Auger Observatory exposure~\cite{PierreAuger:2025eun} for a given UHECR fit model, and the $i$ index runs over the bins above the largest energy data point.

We quantify the preference for a particular model's fit relative to that of another, as follows, at each juncture adopting the most conservative option. First, following PDG practice, as described in the Introduction of~\cite{ParticleDataGroup:2024cfk}, we rescale the fit values of $\chi_\mathrm{i,min}^2$ to $\chi_\mathrm{i,min}^2 (\chi_\mathrm{min,0}^2/N_\mathrm{dof,0})^{-1}$ where the $_0$ indicates the model with the best $\chi_\mathrm{min}^2/N_\mathrm{dof}$.  This is conservative because it reduces the difference between the $\tilde{\chi}_\mathrm{min}^2$'s and hence the strength of the preference which will be inferred.  Next, we account for the difference in the number of degrees of freedom between the fits by taking into account the ``look elsewhere" effect: the model with more parameters allows more flexibility in fitting the data and hence always produces at least as good a $\chi^2$.  Gross and Vitells \cite{GrossVitells10} show that when $N_{\rm dof}$ is large, as is our case where $N_{\rm dof} \geq 55$, an accurate procedure for accounting for the look elsewhere penalty is to define 
\begin{equation}
\tilde{p} \equiv \int_{\Delta\tilde{\chi}^2}^\infty dt \chi^2(t,\Delta_\mathrm{dof}) \ ,
\end{equation}
where $\Delta\tilde{\chi}^2 \equiv \tilde{\chi}^2_\mathrm{1,min} - \tilde{\chi}^2_\mathrm{2,min}$ with $n=\{1,2\}$ labeling the two models being compared, and $\Delta_\mathrm{dof}$ being the absolute difference in the number of degrees of freedom between the two compared models coming from their different number of free fitted parameters. In defining the 2D $\chi^2$ contour levels in plots such as Fig.~\ref{fig:2}, $\Delta_\mathrm{dof} = 2$ since the best fit does not set $f_{56}$ and $f_{80}$ to any particular value. For comparing the fits with and without the jets, $\Delta_\mathrm{dof} = 3$ without processing in the jets, because that is the extra number of parameters describing the jet contribution (spectral index, cutoff, and magnitude). Allowing for processing in the jet, as will be reported elsewhere, would increase $\Delta_\mathrm{dof}$ to 7 on account of the additional parameters $T_{\rm jet}$, $R_{\rm diff,jet}$, $\langle N_{\rm int,jet}\rangle$ and $r_{\rm size,jet}$.

Finally, the number of ``sigma" of preference is~\cite{GrossVitells10,ParticleDataGroup:2024cfk} 
\begin{equation}
Z \equiv \Phi^{-1} (1-\tilde{p}/2) \ ,
\end{equation}
where $\Phi^{-1}$, also known as the quantile function, is the inverse function for the cumulative distribution of a standard Gaussian function.  

The impact of systematic uncertainties in the energy of the primary and $X_{\rm max}$ of the air shower is assessed by repeating the analyses with all combinations of $0,\pm 1\sigma$ systematic shifts, where we take 14\% uncertainty in the energy and 10~g/cm$^2$ for the $X_{\rm max}$. The results are reported in Tables~\ref{tab:jet-p-BNS-shifts1}-\ref{tab:no-jet-SFR-shifts2}.
We do not investigate the impact of the systematic uncertainty in the $X_\mathrm{max}$ scale on $\sigma(X_{\rm max})$ -- the shower-to-shower variation -- since the uncertainty largely cancels out.

The model numbers in the Tables~\ref{tab:jet-p-BNS-shifts1} - \ref{tab:no-jet-SFR-shifts2} correspond to shift combination as follows: 1) $+1\sigma$ energy shift, 2) $+1\sigma$ energy shift and $+1\sigma$ $X_\mathrm{max}$ shift, 3) $+1\sigma$ energy shift and $-1\sigma$ $X_\mathrm{max}$ shift, 4) $+1\sigma$ $X_\mathrm{max}$ shift, 5) $-1\sigma$ $X_\mathrm{max}$ shift, 6) $-1\sigma$ energy shift, 7) $-1\sigma$ energy shift and $+1\sigma$ $X_\mathrm{max}$ shift, and 8) $-1\sigma$ energy shift and $-1\sigma$ $X_\mathrm{max}$ shift.

\newpage

\begin{table} 
\centering
\caption{\textbf{Extra runs with energy shifts and $X_\mathrm{max}$ shifts for the jet-p BNS model.}}
\label{tab:jet-p-BNS-shifts1} 
\begin{tabular}{lcccc} 
\\
\hline
\hline
Parameters & 1 & 2 & 3 & 4 \\
\hline
$\chi^2_\mathrm{tot}/\mathrm{ndf}$ & $99.97/55=1.82$ & $272.65/56 =4.87$ & $157.88/56 = 2.82$ & $301.72/55 = 5.49$ \\
$\lg \langle N_\mathrm{int} \rangle$ & $2.53$ & $2.88$ & $2.42$ & $2.98$ \\
$\gamma_\mathrm{inj}$ & $-2.13$ & $-1.98$ & $-2.27$ & $-1.96$ \\
$\lg (R_\mathrm{max}/\mathrm{V})$ & $18.92$ & $18.90$ & $18.97$ & $18.83$ \\
$\lg (R_\mathrm{diff}/\mathrm{V})$ & $18.04$ & $17.79$ & $17.18$ & $17.68$ \\
$r_\mathrm{size}$ & $7.4$ & $\gtrsim 100$ & $6.2$ & $\gtrsim 100$\\
$T$ [K] & $578$ & $503$ & $536$ & $449$ \\
$\lg (E^\mathrm{jet}/\mathrm{eV})$  & $19.34$ & $19.26$ & $19.43$ & $19.12$ \\
$\gamma^\mathrm{jet}$ & $-1.81$ & $-0.73$ & $-1.98$ & $\approx 0$  \\
$\lg(f^\mathrm{jet})$ & $-1.15$ & $-1.13$ & $-1.10$ & $-1.23$ \\
$f_{1}$ & $\approx 0$  & $\approx 0$  & $\approx 0$  & $\approx 0$ \\
$f_{14}$ & $0.07$ & $0.51$ & $0.11$ & $0.69$ \\
$f_{28}$ & $0.82$ & $0.49$ & $0.46$ & $0.31$ \\
$f_{56}$ & $4.2\times10^{-3}$ & $\approx 0$  & $0.04$ & $\approx 0$  \\
$f_{80}$ & $0.10$ & $\approx 0$  & $0.40$ & $\approx 0$  \\
$f_\mathrm{Gal}$ & $0.67$ & $0.77$ & $0.67$ & $0.74$ \\
$\gamma_\mathrm{Gal}$ & $-2.96$ & $-3.11$ & $-2.94$ & $-3.22$ \\
$\lg( E^\mathrm{Gal}_\mathrm{max}/\mathrm{eV})$ & $18.52$ & $18.69$ & $18.44$ & $18.79$ \\
$f^\mathrm{Gal}_{14}$ & $0.23$ & $0.77$ & $0.01$ & $0.97$ \\
$f^\mathrm{Gal}_{28}$ & $0.77$ & $0.23$ & $0.69$ & $0.03$ \\
$f^\mathrm{Gal}_{56}$ & $\approx 0$ & $\approx 0$  & $0.30$ & $\approx 0$  \\
\hline
\end{tabular}
\end{table}

\begin{table} 
\centering
\caption{\textbf{Extra runs with energy shifts and $X_\mathrm{max}$ shifts for the jet-p BNS model, cont'd.}}
\label{tab:jet-p-BNS-shifts2} 
\begin{tabular}{lcccc} 
\\
\hline
\hline
Parameters & 5 & 6 & 7 & 8 \\
\hline
$\chi^2_\mathrm{tot}/\mathrm{ndf}$  & $157.80/56 = 2.82$ & $139.97/52 = 2.69$ & $391.57/52 = 7.53$ & $148.11/52 = 2.85$ \\
$\lg \langle N_\mathrm{int} \rangle$ & $2.41$ & $2.65$ & $3.24$ & $2.69$ \\
$\gamma_\mathrm{inj}$ & $-2.26$ & $-2.09$ & $-1.99$ & $-2.37$ \\
$\lg (R_\mathrm{max}/\mathrm{V})$ & $18.97$ & $18.79$ & $18.79$ & $18.87$ \\
$\lg (R_\mathrm{diff}/\mathrm{V})$ & $17.26$ & $17.74$ & $17.41$ & $15.97$ \\
$r_\mathrm{size}$ & $4.8$ & $9.9$ & $\gtrsim 100$ & $\gtrsim 100$  \\
$T$ [K] & $544$ & $557$ & $434$ & $575$ \\
$\lg (E^\mathrm{jet}/\mathrm{eV})$  & $19.43$ & $19.03$ & $19.09$ & $19.04$ \\
$\gamma^\mathrm{jet}$ & $-1.97$ &  $\approx 0$ & $\approx 0$ & $\approx 0$ \\
$\lg(f^\mathrm{jet})$ & $-1.10$ & $-1.40$ & $-1.36$ & $-1.22$ \\
$f_{1}$ &  $\approx 0$ &  $\approx 0$ &  $\approx 0$ &  $\approx 0$ \\
$f_{14}$ & $0.10$ & $0.41$ & $0.80$ & $0.05$ \\
$f_{28}$ & $0.48$ & $0.48$ & $0.20$ & $0.63$ \\
$f_{56}$ &  $\approx 0$ & $0.10$ &  $\approx 0$ & $0.04$ \\
$f_{80}$ & $0.41$ &  $\approx 0$ &  $\approx 0$ & $0.28$ \\
$f_\mathrm{Gal}$ & $0.68$ & $0.64$ & $0.69$ & $0.71$ \\
$\gamma_\mathrm{Gal}$ & $-2.95$ & $-3.04$ & $-3.30$ & $-3.22$ \\
$\lg( E^\mathrm{Gal}_\mathrm{max}/\mathrm{eV})$ & $18.43$ & $18.66$ & $18.80$ & $18.61$ \\
$f^\mathrm{Gal}_{14}$ &  $\approx 0$& $0.82$ & $1.00$ &  $\approx 0$ \\
$f^\mathrm{Gal}_{28}$ & $0.70$ & $0.18$ &  $\approx 0$ & $1.00$ \\
$f^\mathrm{Gal}_{56}$ & $0.30$ &  $\approx 0$ &  $\approx 0$ &  $\approx 0$ \\

\hline
\end{tabular}
\end{table}

\begin{table} 
\centering
\caption{\textbf{Extra runs with energy shifts and $X_\mathrm{max}$ shifts for the no-jet SFR model.}}
\label{tab:no-jet-SFR-shifts1}  
\begin{tabular}{lcccc} 
\\
\hline
\hline
Parameters & 1 & 2 & 3 & 4 \\
\hline
\hline
$\chi^2_\mathrm{tot}/\mathrm{ndf}$ & $421.77/59 = 7.15$ & $1074.64/59 = 18.21$ & $531.34/59 = 9.01$ & $959.69/58 = 16.55$ \\
$\lg \langle N_\mathrm{int} \rangle$ & $2.26$ & $2.62$ & $0.84$ & $2.33$ \\
$\gamma_\mathrm{inj}$ & $-1.01$ & $-1.00$ & $-1.14$ & $-1.05$ \\
$\lg (R_\mathrm{max}/\mathrm{V})$ & $18.83$ & $18.85$ & $18.87$ & $18.79$ \\
$\lg (R_\mathrm{diff}/\mathrm{V})$ & $15.83$ & $16.89$ & $15.45$ & $18.01$ \\
$r_\mathrm{size}$ & $\gtrsim 100$ & $\gtrsim 100$ & $\gtrsim 100$ & $\gtrsim 100$ \\
$T$ [K] & $3998$ & $6875$ & $600$ & $8992$ \\
$f_{1}$ & $0.48$ & $0.44$ & $0.29$ & $0.29$ \\
$f_{14}$ & $0.42$ & $0.51$ & $0.65$ & $0.67$ \\
$f_{28}$ & $0.09$ & $0.05$ & $\approx 0$ & $0.04$ \\
$f_{56}$ & $\approx 0$ & $\approx 0$ & $\approx 0$ & $\approx 0$ \\
$f_{80}$ & $0.01$ & $\approx 0$ & $0.05$ & $\approx 0$ \\
$f_\mathrm{Gal}$ & $0.88$ & $0.86$ & $0.89$ & $0.81$ \\
$\gamma_\mathrm{Gal}$ & $-3.19$ & $-3.14$ & $-2.59$ & $-3.15$ \\
$\lg(E^\mathrm{Gal}_\mathrm{max}/\mathrm{eV})$ & $18.57$ & $18.51$ & $18.13$ & $18.48$ \\
$f^\mathrm{Gal}_{14}$ & $0.37$ & $0.59$ & $0.72$ & $0.75$ \\
$f^\mathrm{Gal}_{28}$ & $0.50$ & $0.41$ & $\approx 0$ & $0.25$ \\
$f^\mathrm{Gal}_{56}$ & $0.13$ & $\approx 0$& $0.28$ & $\approx 0$ \\ 
\hline
\end{tabular}
\end{table}

\begin{table} 
\centering
\caption{\textbf{Extra runs with energy shifts and $X_\mathrm{max}$ shifts for the no-jet SFR model, cont'd.}}
\label{tab:no-jet-SFR-shifts2}  
\begin{tabular}{lcccc} 
\\
\hline
\hline
Parameters & 5 & 6 & 7 & 8 \\
\hline
\hline
$\chi^2/\mathrm{ndf}$ & $454.03/58 = 7.83$ & $301.06/55 = 5.47$ & $892.85/55 = 16.23$ & $389.42/55 = 7.08$ \\
$\lg \langle N_\mathrm{int} \rangle$ & $1.36$ & $2.68$ & $3.05$ & $2.30$ \\
$\gamma_\mathrm{inj}$ & $-1.21$ & $-1.53$ & $-1.54$ & $-1.62$ \\
$\lg (R_\mathrm{max}/\mathrm{V})$ & $18.81$ & $18.78$ & $18.79$ & $18.78$ \\
$\lg (R_\mathrm{diff}/\mathrm{V})$ & $16.37$ & $15.80$ & $17.34$ & $15.39$ \\
$r_\mathrm{size}$ & $67$ & $\gtrsim 100$& $\gtrsim 100$& $\gtrsim 100$ \\
$T$ [K] & $900$ & $903$ & $3983$ & $1037$ \\
$f_{1}$ & $0.21$ & $\approx 0$ & $\approx 0$ & $0.12$ \\
$f_{14}$ & $0.69$ & $0.84$ & $0.98$ & $0.46$ \\
$f_{28}$ & $0.04$ & $0.10$ & $0.02$ & $0.31$ \\
$f_{56}$ & $\approx 0$ & $0.07$ & $\approx 0$ & $\approx 0$\\
$f_{80}$ & $0.06$ & $\approx 0$ & $\approx 0$ & $0.11$ \\
$f_\mathrm{Gal}$ & $0.86$ & $0.78$ & $0.70$ & $0.79$ \\
$\gamma_\mathrm{Gal}$ & $-3.08$ & $-3.27$ & $-3.18$ & $-3.18$ \\
$\lg( E^\mathrm{Gal}_\mathrm{max}/\mathrm{eV})$ & $18.41$ & $18.58$ & $18.49$ & $18.46$ \\
$f^\mathrm{Gal}_{14}$ & $0.28$ & $0.53$ & $0.78$ & $\approx 0$ \\
$f^\mathrm{Gal}_{28}$ & $0.51$ & $0.47$ & $0.22$ & $0.92$ \\
$f^\mathrm{Gal}_{56}$ & $0.21$ & $\approx 0$ & $\approx 0$ & $0.08$ \\

\hline
\end{tabular}
\end{table}

\clearpage 

\end{document}